\documentclass[12pt]{article}
\usepackage{graphicx}
\usepackage{amssymb}
\usepackage{amsfonts}

\begin{document}
\title{Dominance for Containment Problems} 
\author{Waseem Akram and
	Sanjeev Saxena\\
Dept. of Computer Science and Engineering, \\
	Indian Institute of Technology, Kanpur, INDIA-208 016\\
	Email:{\{akram,ssax\}@iitk.ac.in}}

\maketitle

\begin{abstract}
In a containment problem, the goal is to preprocess a set of geometric
objects so that, given a geometric query object, we can report all the
objects containing the query object. We consider the containment problem where input objects are homothetic triangles and the query objects considered are line segments, circles, and trapezoids with bases parallel to either axis. We show that this problem can be solved using the $3$-d query dominance problem. The solutions presented can also be extended for higher dimensions.
\end{abstract}


\newpage

\section{Introduction} 

In $d$-dimensional space, a point $p$ dominates another point $q$,
denoted by ``$q \preceq p$'', if each coordinate of $p$ is equal to or
larger than the corresponding coordinate of $q$ \cite{PREPARATA85}. 
The relation $\preceq$ is called dominance relation; the dominance relation is a partial order (for $d>1$). In the dominance reporting problem, the goal is to preprocess a given set of points in $\mathbb{R}^d$ so that we can quickly report all the points that are dominated by a query point \cite{SHI04}. The problem is one of the well-studied problems in computational geometry and has
numerous applications \cite{CHAZELLE86,PREPARATA85,SHI04}. 

In $d$-dimensional space, an object $O_1$ contains another object
$O_2$ if every point of $O_2$ is also a point of $O_1$.  In a
containment searching problem, one wishes to find a data structure for a
set of geometric objects such that, for given a query object
$q$, we can find all objects $p$ that contain $q$. The containment
problem where both input and query objects are axes-parallel
rectangles is known as the rectangle containment problem. It is one of the
most-studied problems of  this kind \cite{CHAZELLE86,EDELSBRUNNER82}. 

The containment problem and dominance problem are related to each
other \cite{AFSHANI14,EDELSBRUNNER82,GUPTA95,OVERMARS81}. Overmars et
al. \cite{OVERMARS81} showed that the following three geometric
searching problems are equivalent: the rectangle enclosure problem,
the rectangle containment problem, and the dominance problem.
Edelsbrunner and Overmars \cite{EDELSBRUNNER82} have shown the
equivalence of the rectangle enclosure/containment/intersection
problems and the dominance problem.

A set of polygons is said to be homothetic if each polygon can be
obtained from any other polygon in the family using scaling and translating
operations. In this paper, we investigate the containment problem where we have to preprocess a set of homothetic triangles such that we can find all the triangles containing a query object. The query objects considered are line
segments, rectangles, circles, ellipses and trapezoids. We show that
after preprocessing a set of homothetic triangles, we can answer a containment query with any of the objects mentioned above using $3$-d dominance queries.
 
Moreover, our solution for the problem, with points as query objects, can be extended for higher dimensions. More precisely, the problem of finding homothetic simplexes in $\mathbb{R}^d$ containing a query point can be solved using $(d+1)$-dimensional dominance queries.

In Section~2, we first show that the interval overlapping problem can
be solved using the $2$-d dominance query problem. Section~3
considers the special case when input triangles are isosceles
right-angled triangles. Arbitrary right-angled triangles are
considered in Section~4 and general triangles in Section~5. 
Section~6 includes extension of the solution for higher dimensions.
We conclude our work in Section~7.

\section{The Interval Overlapping Problem}

The one-dimensional point enclosure problem is a well-studied problem
\cite{CHAZELLE86} where the goal is to preprocess a set of possibly
overlapping intervals on the real line so that we can report all
intervals containing (or overlapping) a query point.  This problem can
easily be transformed to the $2$-d dominance query problem.
Let $I$ be a set of $n$ (possibly overlapping) intervals on the real
line. An interval $[a_i, b_i]\in I$ will contain a point $q$ on the line if and
only if $a_i \le q \le b_i$. In other words, the point $(q, -q)$
should dominate the point $(a_i, -b_i)$. We create a set $S = \{(a_i, -b_i) : [a_i, b_i])\in I\}$ of points in the plane and associate each interval in $I$ with the corresponding created point in $S$.
After preprocessing the set $S$ into a $2$-dominance query structure \cite{SHI04}, denoted by $D_2$, we can find all the intervals containing a query point $q$ by querying the structure $D_2$ with point $(q, -q)$. The intervals associated with the returned points are exactly those which contain the query point $q$.

A more general containment problem where query objects are intervals 
can also be solved using the $2$-d dominance queries. Let $[x_1, x_2]$
be an interval on the line. An interval $[a_i, b_i]\in I$ would
contain the interval $[x_1, x_2]$ if and only if $a_i \le x_1$ and
$x_2 \le b_i$. In other words, the point $(x_1, -x_2)$ should dominate
the point $(a_i, -b_i)$. Note that we can use the same structure $D_2$
built above to find all the points dominated by $(x_1, -x_2)$, and
hence the intervals containing all the query segment $[x_1, x_2]$.

This is a particular instance of the result by Edelsbrunner
\cite{EDELSBRUNNER82}: the rectangle containment problem in $\mathbb{R}^d$ can
be solved using the dominance query problem in $\mathbb{R}^{2d}$.

Remark: For the problem where one is interested in finding the
intervals that are contained in query interval, we preprocess the points $(-a_i, b_i)$, for each $[a_i, b_i]\in I$, for the $2$-d dominance queries. For a query interval $[x_1,x_2]$, we make a $2$-d dominance query with point $(-x_1, x_2)$ and report all the intervals associated with the returned points.

Let us now consider the interval overlapping problem where the goal is
to compute all intervals $p\in I$ intersecting a query interval $q$ i.e. $p\cap q\ne \phi$. An interval $[a_i, b_i]\in I$ intersects an interval $[x_1, x_2]$ if interval $[a_i, b_i]$ contains an endpoint of $[x_1, x_2]$ or $[a_i, b_i]$ is contained in $[x_1, x_2]$. So in order to compute all the intervals intersecting the interval $[x_1, x_2]$, the following $2$-dominance queries (to the structure $D_2$) are needed.
\begin{itemize}
	\item for intervals containing endpoint $x_j$, query with point $(x_j, -x_j), j =1,2$
	\item for intervals containing the whole interval $[x_1, x_2]$, query with $(x_1, -x_2)$
\end{itemize}
{\textbf{Theorem~1:}}
The one-dimensional point enclosure problem, the interval containment
problem, and the interval overlapping problem can be solved using a
constant number of the $2$-d dominance queries.

As the $2$-d dominance query problem can be solved in $O(\lg n+k)$ time, where
$k$ is the number of objects reported \cite{MARKIS98,SHI04,SAXENA09}
after $O(n\log n)$ preprocessing and $O(n)$ space, the above theorem
achieves the same bounds as \cite{CHAZELLE86}.

\section{Containment in Isosceles Right-angled Triangles}

In this section, we consider a containment problem where the input objects are isosceles right-angled triangles and query objects are line segments, rectangles, circles, ellipses and trapezoids.
We will show that after preprocessing the input triangles, each
containment query (with any of the objects mentioned earlier) can be answered by employing a $3$-d dominance query.

Let $S = \{T_1, T_2, ..., T_n\}$ be a set of $n$ isosceles right-angled
triangles. Without loss of generality, we assume that their equal sides parallel to the axes and the right-angled vertices are at the bottom-left position (i.e. minimum $x$ and minimum $y$ coordinates). 
Let triangle $T_i \in S$, for each $i\in [1,n]$, be identified using the coordinates $(a_i, b_i)$ of its right-angled vertex and the length $\alpha_i$ of its equal sides. Note that a point $(h, k)$ will lie below or on (the line through) the hypotenuse of a triangle $T_i\in S$ iff $h+k \le a_i+b_i+\alpha_i$.

For each triangle $T_i\in S$, we create a $3$-d point $(a_i, b_i,
-a_i-b_i-\alpha_i)$ and associate the triangle $T_i$ with it. We then
preprocess these points for the $3$-d dominance queries. We denote the $3$-d dominance query structure by $D_3$.

\subsection{Line Segments}

We partition the query space into three categories: orthogonal
(horizontal or vertical) segments, segments with positive slope and
segments with negative slope. We deal with each of them separately. A
triangle $T_i\in S$ will contain a line segment with endpoints $(x_1, y_1)$
and $(x_2, y_2)$ iff both the endpoints lie inside the triangle (as a
triangle is a convex object). Let $PQ$ be a line segment in the plane with endpoints $P(x_1, y_1)$ and $Q(x_2, y_2)$ such that $x_1 \le x_2$.

\subsubsection{Orthogonal Segments}

We first consider the case when the segment $PQ$ is horizontal i.e. $y_1 = y_2 = y_0$ (say). Observe that a triangle $T_i\in S$ will contain the horizontal segment $[x_1, x_2]\times y_0$ iff the left endpoint of the segment lies inside
$T_i$, and the right endpoint lies below the hypotenuse of triangle $T_i$.

A point inside the triangle $T_i$ will always dominate the
right-angled vertex $(a_i, b_i)$. A point would lie below the line
through the hypotenuse of $T_i$ only if the sum of the point
coordinates is less than or equal to ($a_i+b_i+\alpha_i$). Therefore,
the triangle $T_i$ will contain the segment only if
\begin{itemize}
	\item point $(x_1, y_0)$ dominates the point $(a_i, b_i)$, and
	\item the value $x_2+y_0$ is not more than ($a_i+b_i+\alpha_i$)
\end{itemize}
Thus, the triangle $T_i$ will contain the segment $[x_1, x_2]\times
y_0$ only if the point $(a_i, b_i, -a_i-b_i-\alpha_i)$ is dominated by
the point $(x_1, y_0, -x_2-y_0)$.
\begin{figure}[h]
	\centering
	\includegraphics[scale=2]{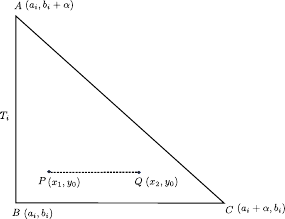}
	\caption{Triangle $T_i$ containing horizontal segment $PQ$.}
\end{figure}

For a horizontal query segment $[x_1, x_2]\times y_0$, we query the
structure $D_3$ with the point $(x_1, y_0, -x_2-y_0)$. We report the
triangles associated with the points returned by $D_3$ in response to the query.

Let us now consider the case when the line segment $PQ$ is a vertical segment i.e. $x_1 = x_2 = x_0$ (say). A triangle $T_i\in S$ will contain the segment $x_0 \times [y_1, y_2]$ if point $(x_0, y_1)$ dominates $(a_i, b_i)$ and point $(x_0, y_2)$ lies below the hypotenuse. So, we can find all triangles of $S$ containing a vertical segment $x_0 \times [y_1, y_2]$ by querying the structure
$D_3$ with point $(x_0, y_1, -x_0-y_2)$.

\subsubsection{Segments with Positive Slope}

Segment $PQ$ with a positive slope will always have $y_1 < y_2$ as $x_1 \le x_2$ (by assumption), where $(x_1,y_1)$ and $(x_2,y_2)$ are coordinates of point $P$ and $Q$ respectively. Note that the endpoint $Q$ is dominating the
endpoint $P$. As a consequence, in order to check whether endpoints
$P$ and $Q$ are inside a triangle $T_i\in S$, it is sufficient to check that $(i)$ $P$
is dominating the point $(a_i, b_i)$, and $(ii)$ $Q$ is below the line
through the hypotenuse.

The triangle $T_i$ will contain segment $PQ$ if $x_1 \ge a_i,$ $ 
y_1 \ge b_i$, and $-x_2-y_2 \ge -a_i-b_i-\alpha_i$.
\begin{figure}[h]
	\centering
	\includegraphics[scale=2]{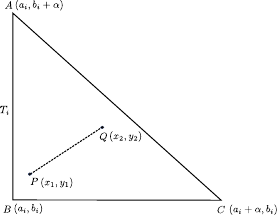}
	\caption{Triangle $T_i$ containing segment $PQ$ with a positive slope.}
\end{figure}
Interpreting it in $3$-d space, the triangle $T_i$ will contain the
segment $PQ$ if the point $(x_1, y_1, -x_2-y_2)$ dominates the point
$(a_i, b_i, -a_i-b_i-\alpha_i)$. We can use the same structure $D_3$
computed above. Thus, we find all the triangles in $S$ containing the segment $PQ$ by querying the structure $D_3$ with point $(x_1, y_1, -x_2-y_2)$.

\subsubsection{Segments with Negative Slope}

Let the segment $PQ$ has a negative slope. Since $x_1 \le x_2$ (by assumption) , so $y_1 > y_2$. We use the following lemma to categorise the queries in two categories.

{\textbf{Lemma~1:}}
The endpoints $P$ and $Q$ will lie inside a triangle $T_i\in S$ if
$y_2\ge b_i$, $x_1\ge a_i$ and $\max\{x_1+y_1, x_2+y_2\}$ is less than
or equal to $a_i + b_i +\alpha_i$.

Proof:
By assumption, we have $x_1 \le x_2$. Since $PQ$ has a negative slope,
then $y_2 < y_1$. The endpoints $P$ and $Q$ will dominate $(a_i, b_i)$
if $x_1 \ge a_i$ and $y_2 \ge b_i$. The third given condition
$\max\{x_1+y_1, x_2+y_2\}\le a_i + b_i +\alpha_i$ makes sure that $P$
and $Q$ (hence the complete segment) are lying below the line through the hypotenuse. 
$\square$

If the slope of the segment $PQ$ is more (less) than that of the
hypotenuses of the triangles in $S$, then the value $x_2+y_2$ will be greater (less) than the
value $x_1+y_1$. 
\begin{figure}[h]
	\centering
	\includegraphics[scale=2]{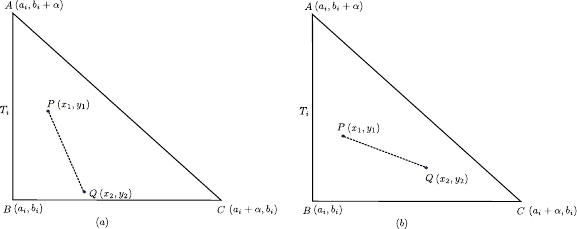}
	\caption{Triangle $T_i$ containing segment $PQ$ with negative slopes. In $(a)$, slope of $PQ$ is smaller than that of hypotenuse's while in $(b)$ $PQ$'s slope is larger than that of hypotenuse.}
\end{figure}
Therefore, a triangle $T_i$ will contain the segment $PQ$ with slope
in range $(0, -1]$ if $x_1 \ge a_i$,{ } $y_2 \ge b_i$, { and
} $-x_2-y_2 \ge -a_i-b_i-\alpha_i$. For the other case when $PQ$ has
slope less than that of the hypotenuse, the triangle $T_i$ will
contain the segment $PQ$ if $x_1 \ge a_i$,{ } $y_2 \ge b_i$, {
and } $-x_1-y_1 \ge -a_i-b_i-\alpha_i$ Thus, we can find all the
triangles containing a line segment with a negative slope by using a
$3$-d dominance query to the data structure $D_3$.

\subsection{Axes-parallel Rectangle}

Let $R$ be an axes-parallel rectangle in the plane with $(x_1, y_1)$
and $(x_2, y_2)$ as coordinates of the extreme endpoints such that
$x_1<x_2$ and $y_1<y_2$. A triangle $T_i\in S$ will contain $R$ iff 
$a_i<x_1, b_i<y_1, (x_2+y_2) < (a_i+b_i+\alpha_i)$. Therefore, using a
$3$-d dominance query $(x_1,y_1,-x_2-y_2)$ to $D_3$, we can find all triangles
containing an axes-parallel rectangle.

Remark: Triangle contains rectangle, iff it contains its diagonals.
Thus, we can also use the method for the case when query is a line segment.

\subsection{Ellipse} 

Let $E$ be an ellipse in the plane. Let $(x_1,y_1)$ (resp. $(x_2,y_2)$) be the coordinates of the point on its boundary with the minimum $y$-coordinate (resp. $x$-coordinate), see fig.~4. 
The ellipse $E$ will be contained in a triangle $T_i\in S$ if all its
boundary points lie inside the triangle.
\begin{figure}[h]
	\centering
	\includegraphics[scale=2]{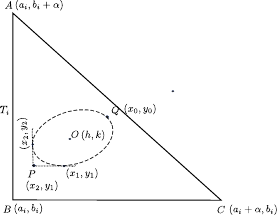}
	\caption{Triangle $T_i$ containing circle $C$.}
\end{figure}
We know that there will be exactly two tangents on the ellipse
parallel to the hypotenuse of $T_i$. Let $Q (x_0, y_0)$ be the tangent
point with the larger coordinate sum.  Observe that the triangle $T_i$ will contain the ellipse $E$ if $a_i \le x_2, b_i \le y_1$ and the line through its hypotenuse lies above the point $(x_0, y_0)$ i.e. $x_0 + y_0 \le a_i+ b_i +\alpha_i$. We arrive at a $3$-d dominance query. Therefore, we can find all the triangles containing $E$ by querying the structure $D_3$ with point $(x_2, y_1, -x_0-y_0)$.

For the particular case of circle, the point $Q$ will have coordinates
$(h + \frac{r}{\sqrt{2}}, k + \frac{r}{\sqrt{2}})$ where $(h, k)$ is
the centre of the circle and $r$ is its radius. The lowest and
leftmost points on the circle will be $(h, k-r)$ and $(h-r, k)$. Thus,
the triangles $T_i\in S$ such that $a_i\le h-r, b_i \le k-r$, {
and }$ h + k + \sqrt{2} \le a_i + b_i + \alpha_i$ will contain the
circle.

\subsection{Trapezoids}

Let $PQRS$ be a trapezoid with bases $PQ$ { and } $SR$ parallel to
the $x$-axis, see fig. $5$. Let $(x_1, y_1), (x_2, y_1), (x_3, y_2)$ and $(x_4, y_2)$ be the coordinates of vertices $P, Q, R,$ and $S$ respectively.
Without loss of generality, we assume that the vertex with the minimum $x$-coordinate is on the lower base. The other case can be handled analogously.

\begin{figure}[h]
	\centering
	\includegraphics[scale=2]{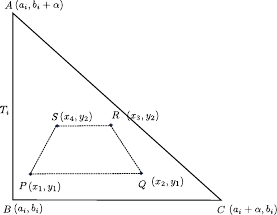}
	\caption{Triangle $T_i$ containing trapezoid $PQRS$ with bases parallel to the $x$-axis.}
\end{figure}

As both triangle and trapezoid are convex geometric objects, the triangle $T_i\in S$ will contain the trapezoid iff all the four vertices of the trapezoid are inside the triangle. Observe that the triangle $T_i$ will contain a point $(p_x, p_y)$ in the plane if the point $(p_x, p_y)$ is dominating the point $(a_i, b_i)$ and the line through its hypotenuse lies above the point $(p_x, p_y)$. By the assumption, vertices $Q, R,$ and $S$ are dominating vertex $P$. Since the dominance is a partial order relation, so if $P$ dominates the point $(a_i, b_i)$, then all other vertices will also dominate it. 

A point lies below the line through the hypotenuse if the sum of its
coordinates is no more than the value $a_i + b_i+\alpha_i$. 
If the vertices $Q(x_2, y_1)$ and $R(x_3, y_2)$ are lying below the line through the hypotenuse, then all other vertices will also lie below it as the sum of coordinates of each vertex is at most $\max\{x_2+y_1, x_3+y_2\}$.
Therefore, all the triangles in $S$ containing the trapezoid in the
plane can be found using a $3$-d dominance query with point $\left(x_1,y_1,\min(-x_3-y_2,-x_2-y_1)\right)$.\\

We combine the results obtained in this section in the following lemma. 
\textbf{Lemma~2:}
A set of isosceles right-angled triangles can be preprocessed such
that, given a query object (line-segments, rectangles, circles,
ellipses, trapezoids), we can find all triangles containing the query
object by employing a $3$-d dominance query.

Remark: We can answer a containment query with any convex object. The
idea is to find the minimum $x$-coordinate and the minimum $y$-coordinate on the object, say $h$ and $k$ respectively and find a point $(x', y')$ on the object with the maximum coordinate sum. The triangles $T_i\in S$
will contain the object if $(h, k)$ dominates $(a_i, b_i)$ and $x'+y'\le a_i+b_i+\alpha_i$.

\section{Arbitrary Right-angled Triangles}

In this section, we consider the case when the homothetic triangles in $S$ are right-angled (not necessarily isosceles).  Let the input triangles have horizontal and vertical sides parallel to the axes and their right-angled vertices are at the bottom-left position. If not, we can achieve it using a rotation operation. By a scaling operation, a right-angled triangle can be transformed into an isosceles right-angled triangle. Under translating and scaling operations, each query object (line segment, trapezoid, ellipse) transforms into another object of the same type.

Let the hypotenuses of triangles in $S$ be parallel to the line
$\frac{x}{a}+\frac{y}{b}=1$. We transform the triangles of $S$ into 
isosceles right-angled triangles by using the scaling transformation
$x = ax'$ { and } $y = by'$. Let $S'$ denote the set of these
transformed triangles.  We preprocess $S'$ for the containment queries
as described in the previous section and denote the structure by
$D_3'$. We can obtain the set $S'$ from the set $S$ in linear time.

For a query object $q$, we first transform it using the scaling
operation $x=ax'$ { and } $y=by'$. Let $q'$ be the transformed
object. We then find all triangles of $S'$ containing $q'$ using
$D_3'$. The triangles from $S$ corresponding to the triangles returned
by $D_3'$ are the ones which contain the query object $q$.

\textbf{Lemma~3:}
The problem of finding all right-angled homothetic triangles
containing a query line-segment, ellipse, or trapezoid can be solved
using $3$-d dominance.

\section{Arbitrary Homothetic Triangles}

We describe the case where homothetic triangles are acute-angled triangles. The other case can be handled analogously. A shear transformation can transform an acute-angled triangle into a right-angled triangle. The objects considered (line segment, ellipse, and trapezoid) do not change their type on a shear transformation.

Without loss of generality, we assume that the triangles in $S$ have a
side parallel to the $x$-axis. Let $m$ be the slope of the left side.
By shear transformation $x' = x -\frac{y}{m}$ { and } $y' = y$, the
triangles in $S$ will transform into right-angled triangles. Let $S'$
be the set of transformed triangles. We then preprocess $S'$ for the
containment queries as explained in the previous section.

Again for a query object $q$, we first transform it into a new object
$q'$ of the same type using the shear transformation.  Observe that
under this transformation, horizontal lines remain horizontal. Thus,
trapezoids will transform to trapezoids. Again ellipses will transform
into ellipses. We then find the triangles from $S'$ containing $q'$.
Finally, we report the triangles of $S$ that correspond to the
returned ones.

\textbf{Theorem~2:}
A set of homothetic triangles in the plane can be preprocessed so that all objects containing a query line segment, ellipse or trapezoid with bases
parallel to either axis can be obtained using $3$-d dominance.

As $3$-dominance queries can be answered in $O(\log n+k)$ time after
$O(n\log n)$ preprocessing time \cite{MARKIS98,SAXENA09}, thus we have
the following theorem.

\textbf{Theorem~3:}
A set of homothetic triangles in the plane can be preprocessed in $O(n\log n)$ time such that all the triangles containing a query line segment, rectangle, circle, ellipse or trapezoid with bases parallel to either axis can be found
in $O(\log n+ k)$ time, where $k$ is the number of triangles containing the query object.

\section{Extension for Higher Dimensions}
If we are to deal with points only as query objects, then our approach of solving the problem using dominance queries can also be extended for higher dimensions. By using parallel arguments, we can show that a set of homothetic simplexes in $\mathbb{R}^d$ can be preprocessed so that, given a point in    $\mathbb{R}^d$, we can report all the simplexes containing the point using a $(d+1)$-dimensional dominance query. For example, in $3$-d space, in place of triangles we will have tetrahedron and we can solve the problem using $4$-dimensional dominance queries.

\section{Conclusion}

In this paper, we considered the problem of finding homothetic triangles containing different objects (line-segment, rectangle, circle, ellipse, or trapezoid with bases parallel to either axis). We have shown that the problems can be solved in optimal $O(\log n+k)$ time; here $k$ is the output size. Pre-processing time is $O(n\log n)$. We also showed that the solution can be extended for higher dimensions in the scenario where query objects are only  points.

Like homothetic triangles, a circle in the plane can be uniquely defined using three parameters. It will be interesting to see whether the dominance
problem can help in finding circles containing query objects like
point, line-segment etc.

\section*{Acknowledgement} 

We would like to thank an anonymous referee for suggesting the use of
dual space for another problem in a different paper.

\bibliographystyle{acm}

\end{document}